\documentclass{iau} 
\usepackage{graphicx}
\usepackage{subfigure}
\usepackage{amsmath}
\usepackage{wrapfig}
\usepackage{hyperref}
\usepackage{gensymb}
\hypersetup{
    colorlinks=true,
    linkcolor=blue,
    filecolor=magenta,
    urlcolor=red,
    citecolor=Magenta,
   pdftitle={Sharelatex Example},
}

\title[Continuous gravitational wave] %% give here short title %%
{Continuous gravitational wave from magnetized white dwarfs}

\author[S. Kalita \& B. Mukhopadhyay]   %% give here short author list %%
{Surajit Kalita$^1$ \and Banibrata Mukhopadhyay$^2$}

\affiliation{Department of Physics, Indian Institute of Science, Bangalore 560012, India \\ $^1$email: {\tt surajitk@iisc.ac.in},
$^2$email: {\tt bm@iisc.ac.in}}

\pubyear{2019}
\volume{357}  %% insert here IAU Symposium No.
\jname{White Dwarfs as probes of fundamental physics and tracers of planetary, stellar \& galactic evolution}
\editors{A.C. Editor, B.D. Editor \& C.E. Editor, eds.}

\begin{document}

\maketitle

\begin{abstract}
Recent evidence of super-Chandrasekhar white dwarfs (WDs), from the observations of over-luminous type Ia supernovae (SNeIa), has been a great astrophysical discovery. However, no such massive WDs have so far been observed directly as their luminosities are generally quite low. Hence it immediately raises the question of whether there is any possibility of detecting them directly. The search for super-Chandrasekhar WDs is very important as SNeIa are used as standard candles in cosmology. In this article, we show that continuous gravitational wave can allow us to detect such super-Chandrasekhar WDs directly.

\end{abstract}

\firstsection % if your document starts with a section, remove some space above using this command.

%==============================================================================================================
\section{Introduction}
Recent detections of gravitational waves (GWs) from LIGO and VIRGO events have opened a new window in astronomy. These GW events were observed when two compact objects in a binary system, such as black holes (BHs) or neutron stars (NSs) or one BH and one NS, merged to produce a single compact object. It is important to note that these events are short lasting. However, there is another type of GW, known as continuous GW, which is continuously emitted with certain amplitude and frequency. We discuss in this article that isolated rotating white dwarfs (WDs) as well as binary WDs are prominent sources producing continuous GW.

During the past decade, a good number of peculiar type Ia supernovae (SNeIa) have been observed, which are over-luminous in nature. It was inferred that these SNeIa must have progenitor WDs with masses $\sim 2.1-2.8 M_\odot$ (\cite[Howell \etal\ 2006]{Howell06}, \cite[Scalzo \etal\ 2010]{Scalzo10}). This immediately indicates the violation of the Chandrasekhar mass-limit (currently accepted value $\sim 1.4 M_\odot$ for non-magnetized static WDs). Many researchers proposed different models to explain these exceedingly massive WDs (e.g. \cite[Das \& Mukhopadhyay 2013]{Das13}, \cite[Kalita \& Mukhopadhyay 2018]{Kalita18}, \cite[Ong 2018]{Ong18}). In this article, we show that in the presence of magnetic fields and rotation, the mass of WDs can increase significantly and eventually they can efficiently emit continuous gravitational radiation. Once such GW is detected by future space-based detectors, it will provide a direct evidence of super-Chandrasekhar WDs.

%==============================================================================================================
\section{Model of the magnetized rotating white dwarfs}

It is well known that GWs have two polarizations, $h_+$ and $h_\times$. Moreover if an object rotates with an angular frequency $\Omega$, with the rotation axis disaligned with the body's principal axis, it will emit gravitational radiation. For such an object, the polarizations at time $t$ are given by (\cite[Zimmermann \& Szedenits 1979]{Zimmermann79}, \cite[Bonazzola \& Gourgoulhon 1996]{Bonazzola96})
\begin{equation}\label{gravitational polarization}
\begin{aligned}
h_+ &= h_0\sin\chi\Bigg[\frac{1}{2}\cos i \sin i\cos\chi\cos\Omega t-\frac{1+\cos^2i}{2}\sin\chi\cos2\Omega t\Bigg],\\
h_\times &= h_0\sin\chi\Bigg[\frac{1}{2}\sin i\cos\chi\sin\Omega t-\cos i\sin\chi\sin2\Omega t\Bigg],
\end{aligned}
\end{equation}
with
\begin{equation}\label{grav_wave_amplitude}
h_0 = -\frac{6G}{c^4}Q_{z'z'}\frac{\Omega^2}{d},
\end{equation}
\begin{wrapfigure}{r}{0.45\textwidth}
  \begin{center}
    \includegraphics[width=0.4\textwidth]{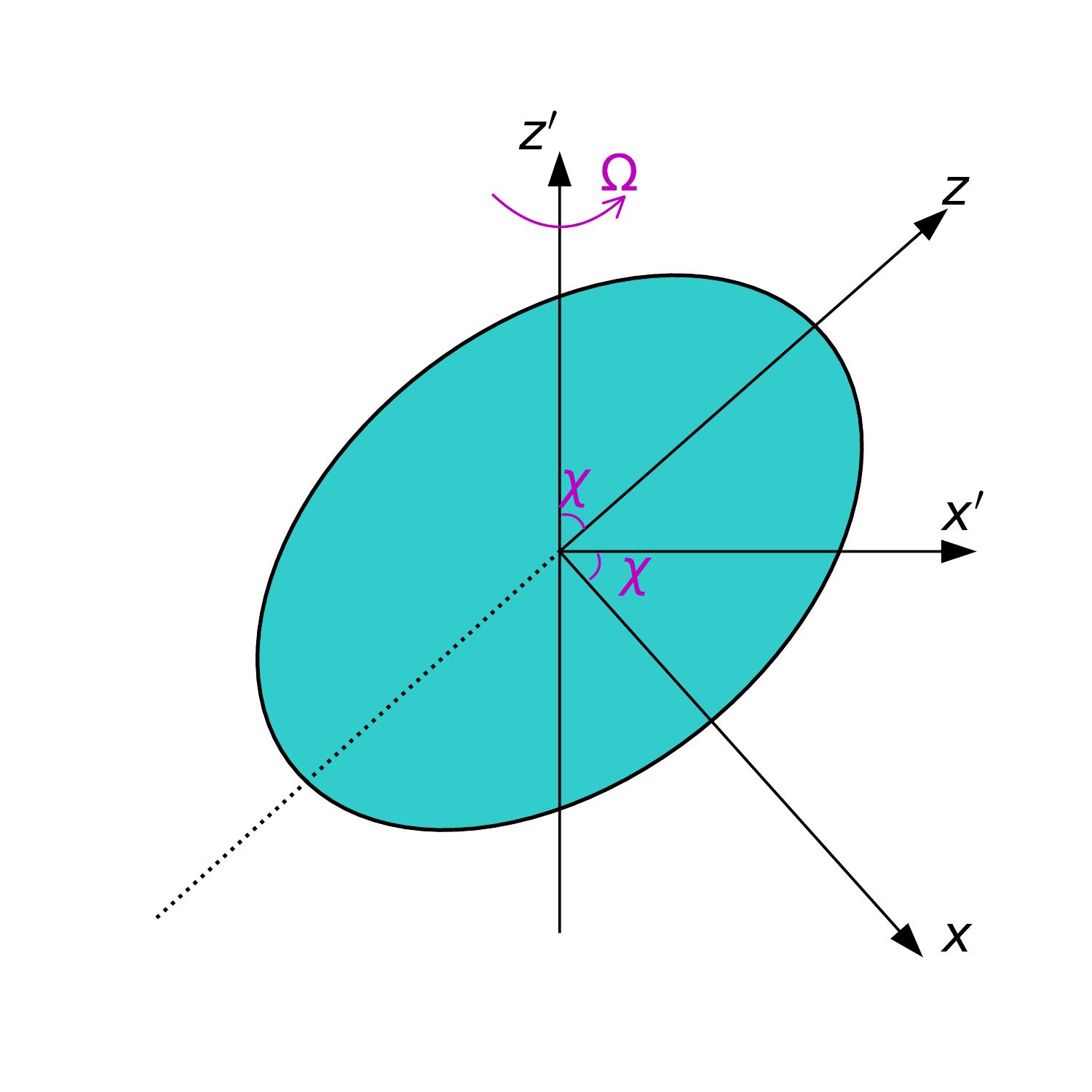}
  \end{center}
\caption{A cartoon diagram of magnetized rotating WD with misalignment between magnetic field and rotation axes.}
\label{Fig: Compact objects}
\end{wrapfigure}
where $Q_{z'z'}$ is the quadrupole moment of the distorted star, $\chi$ the angle between the rotation axis $z'$ and the body's third principal axis $z$, $i$ the angle between the rotation axis of the object and our line of sight, $G$ the Newton's gravitational constant and $c$ the speed of light. It has already been shown that a toroidal magnetic field makes a star prolate, whereas a poloidal magnetic field (as well as rotation) deforms it to an oblate shape (\cite[Subramanian \& Mukhopadhyay 2015]{Subramanian15}). As shown in Figure \ref{Fig: Compact objects}, we assume that the magnetic field is along the $z-$axis, whereas the star is rotating about the $z'-$axis. Using this configuration and from Equation \eqref{grav_wave_amplitude}, the GW amplitude becomes
\begin{equation}
h_0 = \frac{2G}{c^4}\frac{\Omega^2\epsilon I_{xx}}{d}(2\cos^2\chi-\sin^2\chi),
\end{equation}
where $\epsilon = (I_{zz}-I_{xx})/I_{xx}$ is the ellipticity of the body and $I_{xx}$, $I_{yy}$, $I_{zz}$ are the principal moments of inertia of the object about its three principal axes ($x-$, $y-$, $z-$axes respectively). Here we use the {\it XNS} code (\cite[Pili \etal\ 2014]{Pili14}) to model WDs. This code solves the axisymmetric equilibrium configuration of stellar structure in general relativity. A detail discussion about all the parameters used to model WDs is given by \cite[Kalita \& Mukhopadhyay (2019)]{Kalita19}. Moreover, we assume the distance between the WD and the detector to be 100 pc.

%==============================================================================================================
\begin{figure}[htbp]
\centering
\subfigure[Toroidal magnetic field]{\includegraphics[scale=0.37]{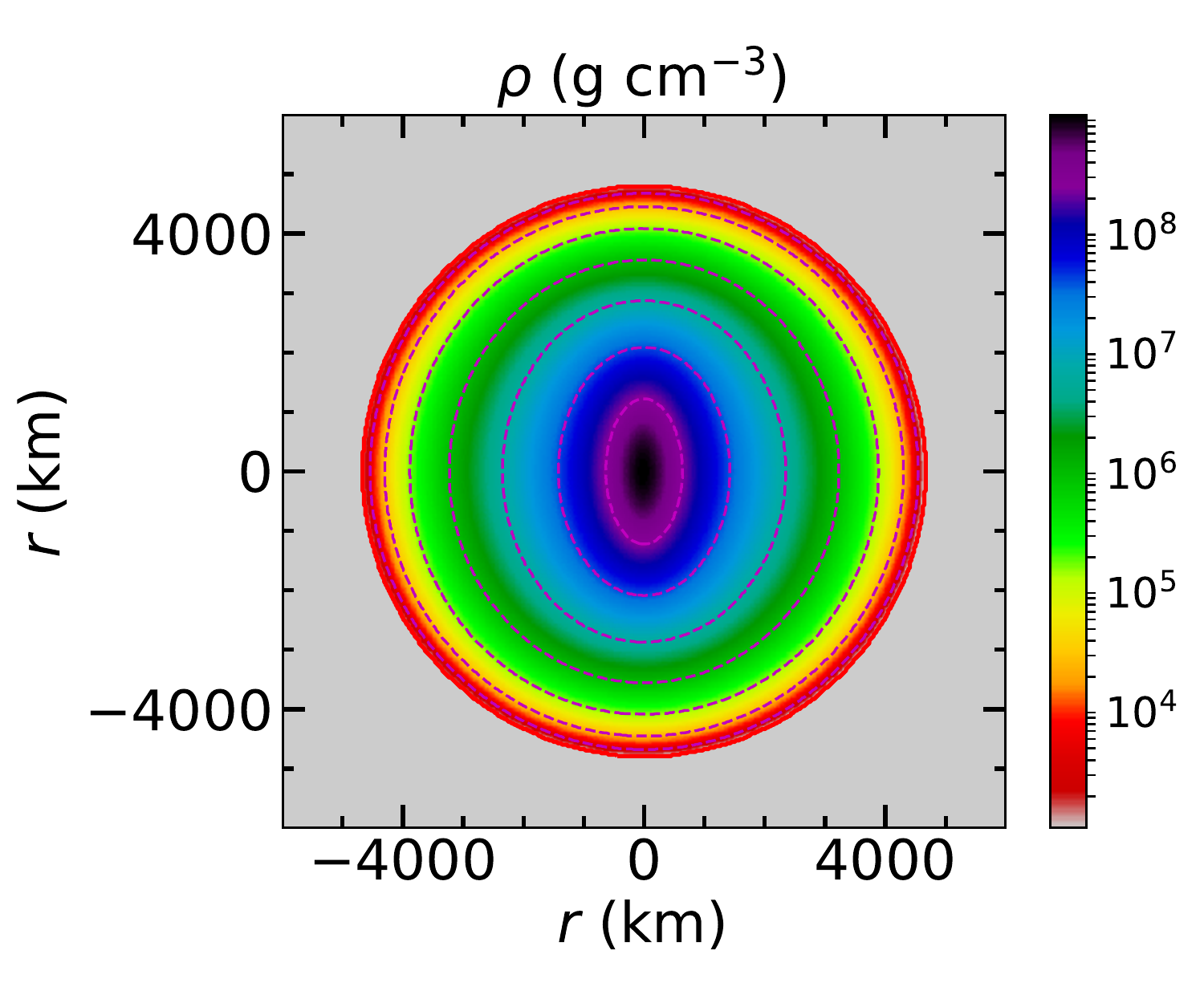}}
\subfigure[Poloidal magnetic field]{\includegraphics[scale=0.37]{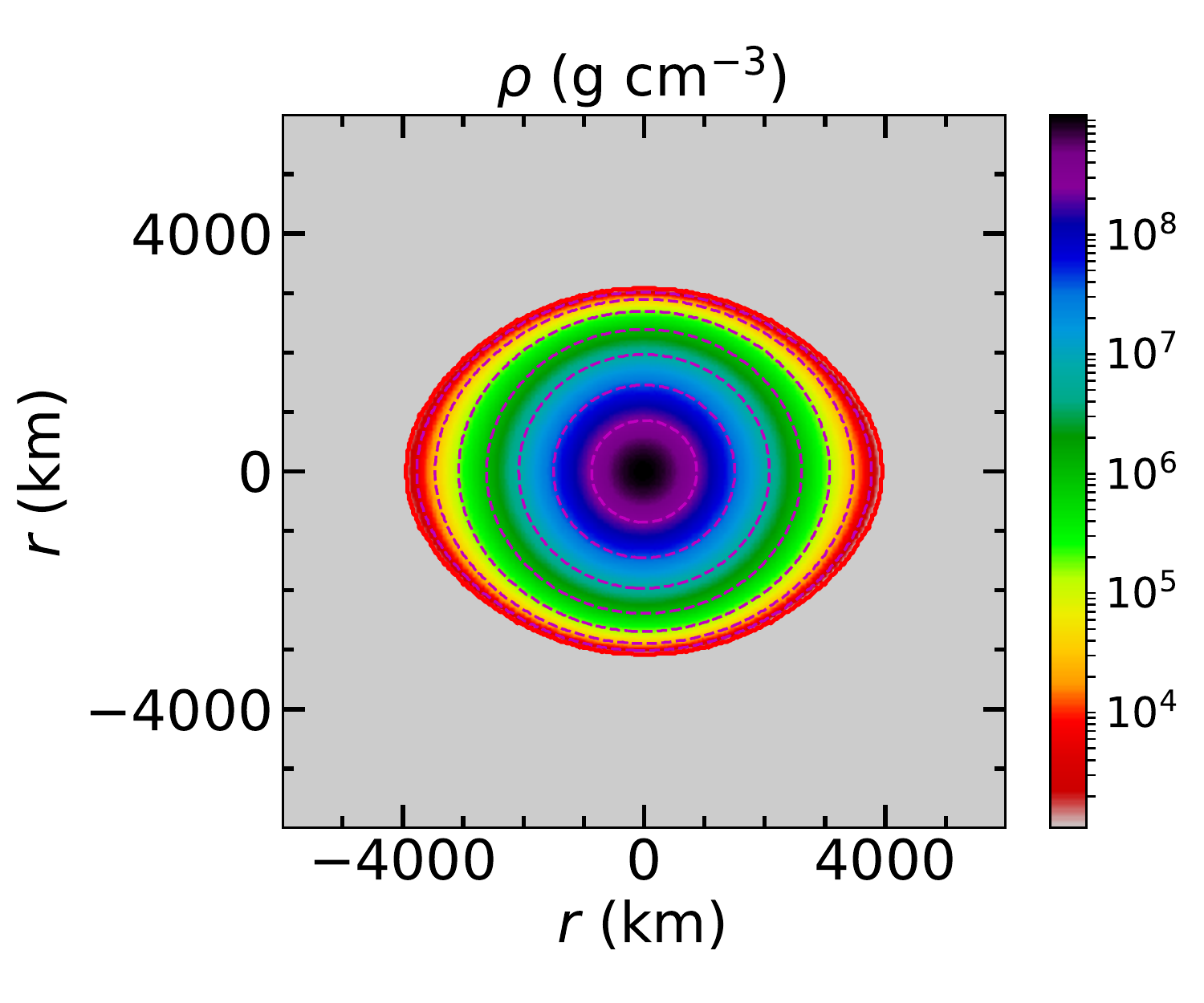}}
\caption{Uniformly rotating B-WDs with (a) $\Omega = 0.02$ rad/s, (b) $\Omega = 1.32$ rad/s. The color code corresponds to density distribution.}
\label{Fig: Magnetized WD}
\end{figure}

\section{Structure of rotating magnetized white dwarfs}

We show typical cases for rotating WDs with toroidal and poloidal magnetic fields in Figure \ref{Fig: Magnetized WD}(a) and \ref{Fig: Magnetized WD}(b) respectively. It is observed that the deformation of the core is more prominent than that of the outer regions. The overall shape of the WDs depends on the combined effects arsing from the rotation and the structure of the magnetic field. It is important to note that the central field could be as large as $\sim 10^{14}$ G even if the surface field is $\sim 10^9$ G (\cite[Heyl 2000]{Heyl00}, \cite[Ferrario \etal\ 2015]{Ferrario15}). The stability of these magnetized WDs (B-WDs) is determined constraining the kinetic to gravitational energy ratio as well as the magnetic to gravitational energy ratio (\cite[Braithwaite 2009]{Braithwaite09}).

%==============================================================================================================
\newpage
\section{Gravitational radiation from rotating magnetized white dwarfs}

\begin{wrapfigure}{r}{0.58\textwidth}
\centering
\includegraphics[scale=0.3]{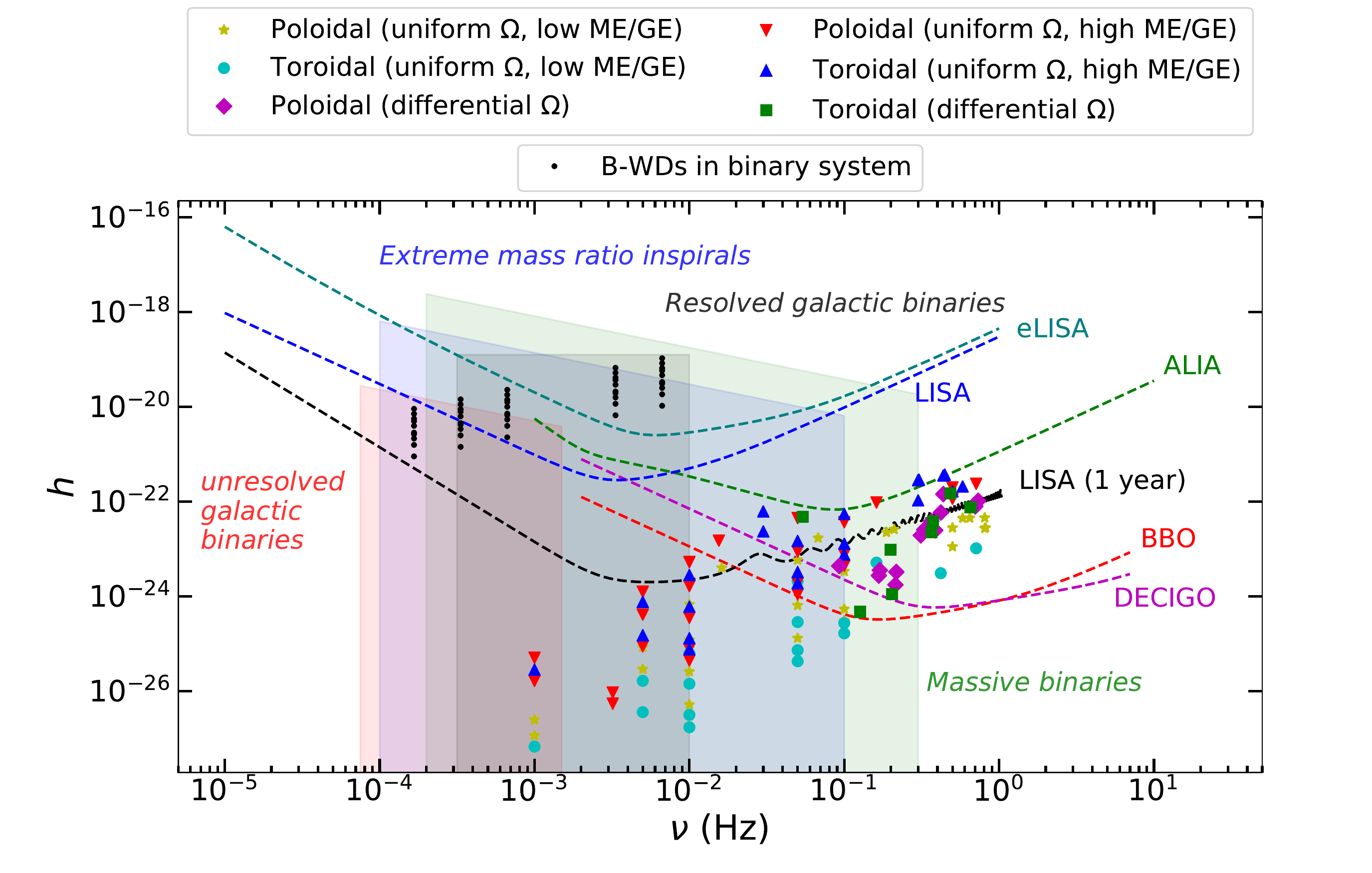}
\caption{Dimensionless GW amplitude for WDs as a function of frequency with $\chi=3\degree$ and $d=100$ pc.}
\label{Fig: Gravitational wave}
\end{wrapfigure}
The gravitational radiation has been calculated for various stable B-WDs by varying the central density and the magnetic field strength and also considering both the uniform and differential rotation. Figure \ref{Fig: Gravitational wave} shows the strength of the GW for these B-WDs assuming the misalignment between the rotation and magnetic field axes to be $3\degree$. It is evident from this figure that many of the isolated rotating B-WDs can be detected by future space-based detectors, such as, LISA, DECIGO, BBO, etc. It is also important to note that these B-WDs cannot be detected by LISA directly, while it might be possible to detect some of them with the help of LISA 1 year integration curve. Moreover, since our galaxy is very rich in binary WDs, they may create some confusion noise (\cite[Ruiter \etal\ 2010]{Ruiter10}). However, since the time period of B-WDs is much smaller than the orbital time period of binaries, there is no problem in distinguishing isolated B-WDs from binaries. Some of the massive black hole binaries may create confusion noise too, which however can easily be removed with proper source modeling, but this is beyond the scope of this work. A similar exploration has also been carried out in the case of neutron stars (\cite[Kalita \& Mukhopadhyay 2018]{Kalita18}).

%==============================================================================================================
\section{Multi-messenger astronomy with magnetized white dwarfs}

Being quadrupolar in nature, GW is associated with quadrupolar luminosity. Moreover, since the B-WDs have magnetic fields as well as rotation, they behave as rotating dipoles, and they must possess luminosity due to dipole radiation along with gravitational radiation, which means B-WDs have electromagnetic counterparts. The luminosity due to electromagnetic dipole radiation is given by
$L_\text{EM} = 4\pi^2I_{z'z'} \dot{P}/P^3,$
where $P$ is the rotational period of the body which is changing with time as $\dot{P}$. Moreover, the GW luminosity is given by
\begin{align}
L_\text{GW} = \frac{G}{5c^5}\left\langle\dddot{Q}_{ij} \dddot{Q}_{ij}\right\rangle 
&= \frac{G\Omega^6}{20c^5} (I_{xx}-I_{zz})^2 \sin^2\chi (2\cos^2\chi - \sin^2\chi)^2 \nonumber\\ &\left\{\cos^2\chi \sin^2 i (1+\cos^2i) + 4\sin^2\chi (1+6 \cos^2i+ \cos^4i) \right\}.
\end{align}
It is clearly observed from the above formulae that the misalignment between the rotation and magnetic field axes is necessary to emit gravitational radiation, whereas this is not compulsory for dipolar radiation. A comparison between the values of $L_\text{GW}$ and $L_\text{EM}$ is given in Table \ref{Luminosity_table}. Since $L_\text{GW}$ is larger than $L_\text{EM}$, it will be another unique way of separating B-WDs from regular WDs.

\begin{table}
\centering
\caption{$L_\text{GW}$ and $L_\text{EM}$ for WDs considering $\dot{P}=10^{-15}$ Hz s$^{-1}$
and $\chi=3\degree$. $M$ and $R$ are respectively the mass and mean radius, $B_s$ is the surface magnetic field at the pole of the B-WDs.}
\label{Luminosity_table}
\begin{tabular}{|l|l|l|l|l|l|l|}
\hline
	$M$ ($M_\odot$) & $R$ (km) & $P$ (s) & $B_s$ (G) & $L_\text{GW}$ (ergs s$^{-1}$) & $L_\text{EM}$ (ergs s$^{-1}$)\\
\hline\hline
	1.420 & 1718.8 & 1.5 & $6.12\times10^{8}$ & $2.91\times10^{35}$ & $5.50\times10^{34}$\\
	1.640 & 1120.7 & 2.0 & $2.78\times10^{9}$ & $3.46\times10^{36}$ & $3.03\times10^{34}$\\
	1.702 & 1027.2 & 3.1 & $4.52\times10^{9}$ & $3.17\times10^{35}$ & $8.23\times10^{33}$\\
	\hline
\end{tabular}
\end{table}

%==============================================================================================================
\section{Conclusion}

Although SNeIa have been used as standard candles, some of them are powered by super-Chandrasekhar WDs which have not been directly detected so far because they are very dim. It is already known that magnetic fields and rotation can increase the mass of WDs. In this article, we have argued that continuous GW is one of the best ways to detect some of these B-WDs directly, provided they posses tri-axial systems. Therefore, if there are some super-Chandrasekhar WDs in the sky, the future space-based GW detectors should be able to detect them.

%==============================================================================================================
\begin{acknowledgement}
S. K. would like to thank IAU for providing the travel grant to attend the symposium. The authors further thank Lilia Ferrario of ANU for thorough reading the manuscript with encouraging amendments.
\end{acknowledgement}

%==============================================================================================================

\end{document}